\documentstyle[12pt]{article}
\newcommand{\journal}[4]{{\em #1~}#2\,(19#3)\,#4;}
\newcommand{\aihp}{\journal {Ann. Inst. Henri Poincar\'e}}

\newcommand{\ijmp}{\journal {Int. J. Mod. Phys.}}

\newcommand{\pr}{\journal {Phys. Rev.}}

\newcommand{\jmp}{\journal {J. Math. Phys.}}

\newcommand{\cmp}{\journal {Comm. Math. Phys.}}

\newcommand{\np}{\journal {Nucl. Phys.}}
\newcommand{\pl}{\journal {Phys. Lett.}}

\newcommand{\prep}{\journal {Phys. Reports}}

\newcommand{\nc}{\journal {Nuovo Cim.}}

\makeatletter
\@addtoreset{equation}{section}
\makeatother

\def\Lp{\displaystyle{\biggl(}}
\def\Rp{\displaystyle{\biggr)}}
\def\LP{\displaystyle{\Biggl(}}
\def\RP{\displaystyle{\Biggr)}}
\newcommand{\lp}{\left(}\newcommand{\rp}{\right)}

\newcommand{\lac}{\left\{}\newcommand{\rac}{\right\}}
\newcommand{\G}{\Gamma}
\newcommand{\D}{\Delta}

\renewcommand{\d}{\delta}

\newcommand{\f}{\phi}
\newcommand{\g}{\gamma}

\newcommand{\m}{\mu}
\newcommand{\n}{\nu}

\renewcommand{\o}{\omega} \renewcommand{\O}{\Omega}

\newcommand{\r}{\rho}
\newcommand{\s}{\sigma} \renewcommand{\S}{\Sigma}

\renewcommand{\AA}{{\cal A}}
\newcommand{\BB}{{\cal B}}
\newcommand{\CC}{{\cal C}}
\newcommand{\DD}{{\cal D}}

\newcommand{\FF}{{\cal F}}
\newcommand{\GG}{{\cal G}}

\newcommand{\LL}{{\cal L}}
\newcommand{\MM}{{\cal M}}
\newcommand{\NN}{{\cal N}}

\newcommand{\QQ}{{\cal Q}}
\newcommand{\RR}{{\cal R}}
\newcommand{\SS}{{\cal S}}
\newcommand{\TT}{{\cal T}}
\newcommand{\UU}{{\cal U}}
\newcommand{\VV}{{\cal V}}

\newcommand{\complex}{{\kern .1em {\raise .47ex
\hbox {$\scriptscriptstyle |$}}
    \kern -.4em {\rm C}}}
\newcommand{\real}{{{\rm I} \kern -.19em {\rm R}}}
\newcommand{\rational}{{\kern .1em {\raise .47ex
\hbox{$\scripscriptstyle |$}}
    \kern -.35em {\rm Q}}}
\renewcommand{\natural}{{\vrule height 1.6ex width
.05em depth 0ex \kern -.35em {\rm N}}}

\newcommand{\tr}{{\rm {Tr} \,}}
\newcommand{\cb}{{\bar c}}
\newcommand{\half}{\frac 1 2}
\newcommand{\pa}{\partial}
\newcommand{\pad}[2]{{\displaystyle{\frac{\partial #1}{\partial #2}}}}
\newcommand{\fud}[2]{{\displaystyle{\frac{\delta #1}{\delta #2}}}}

\newcommand{\ie}{{{\em i.e.}\ }}

\newcommand{\cf}{{\em cf.\ }}
\newcommand{\sla}{\raise.15ex\hbox{$/$}\kern -.57em}
\newcommand{\imply}{\quad\Rightarrow\quad}
\newcommand{\twiddle}{\lower.9ex\rlap{$\kern -.1em\scriptstyle\sim$}}
\newcommand{\afv}{{{\rm all\ fields\ }\varphi}}

\newcommand{\vf}{{\varphi}}

\renewcommand{\=}{&=&} %seems not to work in footnotes
\newcommand{\equ}[1]{(\ref{#1})}
\newcommand{\eq}{\begin{equation}}
\newcommand{\eqn}[1]{\label{#1}\end{equation}}
\newcommand{\eea}{\end{eqnarray}}
\newcommand{\eqa}{\begin{eqnarray}}
\newcommand{\eqan}[1]{\label{#1}\end{eqnarray}}
\newcommand{\ba}{\begin{array}}
\newcommand{\ea}{\end{array}}
\newcommand{\eqac}{\begin{equation}\begin{array}{rcl}}
\newcommand{\eqacn}[1]{\end{array}\label{#1}\end{equation}}

\newcommand{\qq}{&\qquad &}
\renewcommand{\pad}[2]{{\displaystyle{\frac{\partial #1}{\partial #2}}}}

\newcommand{\om}[2]{\omega^{#1}_{#2}}
\newcommand{\ccinq}{c^{(5)}}
\newcommand{\intx}{\int d^4 \! x \, }

\newcommand{\sx}{\SS_\Sigma}
\newcommand{\es}{\\[2mm]}
\newcommand{\Om}[2]{\Omega^{#1}_{#2}}
\newcommand{\uu}[2]{u^{#1}_{#2}}
\newcommand{\ch}[2]{\chi^{#1}_{#2}}
\begin{document}
%********************************************************
\begin{titlepage}
\title{$\ $\\
\vspace{5cm}
{\bf On the finiteness of the BRS modulo--$d$ cocycles} }
\author{
Olivier Piguet and Silvio P. Sorella\thanks{Supported in part by the
Swiss National Science Foundation}\\
{\small D\'ep. de Physique Th\'eorique, Universit\'e de Gen\`eve}\\
 {\small          CH-1211 Gen\`eve 4, Switzerland}\\
}
\maketitle
\thispagestyle{empty}
\begin{center}
{\large{\bf UGVA-DPT 1992/3-759}}
\end{center}
\end{titlepage}
\begin{titlepage}
$\ $\vspace{5cm}
\begin{center}
{\Large{\bf Abstract}}\end{center}
\vspace{12mm}

\noindent
Ladders of field polynomial differential forms obeying
systems of descent equations and corresponding to observables and
anomalies of gauge theories are renormalized. They obey renormalized
descent equations. Moreover they are shown to
have vanishing anomalous dimensions.
As an application a simple proof of the nonrenormalization theorem for
the nonabelian gauge anomaly is given.
\end{titlepage}
%*********************************************************************
\section{Introduction}
A whole class of observables and anomalies
in gauge theories
are built by solving systems of descent
equations~\cite{zumino} for sequences -- ''ladders'' --
of classical field polynomials.
This amounts to studying
the BRS modulo--$d$ cohomolgy in the space of local field
polynomials~\cite{d-violette,brandt}. The nontrivial solutions of the
descent equations are
uniquely specified  by invariant ghost monomials of the general form
$\tr c^n$, $n$ odd,
where $c$ is the usual Faddeev-Popov ghost belonging to the
Lie algebra of the gauge group.

The aim of the present paper is twofold. First, show the perturbative
existence and uniqueness of quantum insertions, for any given classical
ladder, which fulfil the quantum version of the descent
equations. Second, prove that these quantum insertions are
''ultraviolet finite'', \ie that they  are
characterized by vanishing anomalous
dimensions~\cite{bms,lps}.

The proof of ultraviolet finiteness will be given
in a particular gauge, namely the Landau gauge~\cite{bps-ghost}.
However, thanks to the gauge invariance of the ghost cocycle $\tr c^n$
and by means of the extended BRS technique~\cite{ps-extbrs}, its
validity persists in a generic covariant gauge, as shown for a
particular case in~\cite{lps}.
The choice of the Landau
gauge is the natural one for studying the ultraviolet properties of such
ghost monomials. Indeed, as shown in~\cite{bps-ghost},
the Landau gauge    is characterized by the ''ghost equation'' which
controls the dependence of the theory on the ghost field $c$. It is
this equation which implies the vanishing of the anomalous dimension of
$c$ and also of all cocycles $\tr c^n$.

These results were already proved, by using Feynman graph considerations,
in the case of the ladder related to the
cocycle $\tr c^3$: this was a basic ingredient
in the proof of the nonrenormalization theorem of the U(1) anomaly~\cite{lps}
in renormalizable nonabelian gauge theories.

We will give here the proof refering to the ladder
$n=5$ for simplicity, but in a way suitable for generalization to any
$n$. This case is of particular importance for gauge theories in
4-dimensional space-time due to the relation of this ladder with the
nonabelian gauge anomaly. As an application of our results we present
here a completely algebraic proof of the nonrenormalization
theorem~\cite{bardeen,padua,bbbc} for the latter anomaly.

The paper is organized as follows. In Section 2 we introduce the cocycle
$\tr c^5$ in the framework of a general renormalizable gauge theory,
and we write down all the functional identities characterizing the
properties of the model in the classical approximation. Section 3 is
devoted to the renormalization of that cocycle. The ultraviolet
finiteness of the renormalized $\tr c^5$ is proved
in Section 4 by showing that it obeys a Callan-Symanzik equation without
anomalous dimension. Section 5 extends the proof to the whole renormalized
$n=5$ ladder, and in particular to the anomaly insertion. In Section 6
we present the algebraic proof of the nonrenormalization theorem. A
useful proposition on local cohomology is given in the Appendix.
%*************************************************************************
\section{The $\ccinq$ cocycle}\label{section2}  %2
\subsection{The functional identities}\label{identities}
We consider here a massless gauge theory in 4-dimensional space-time.
The gauge group is a compact Lie group $G$, assumed to be simple.
Its generators obey the commutation rules
\eq
[T_a,T_b] = i f_{abc}T_c   \ .
\eqn{lie}
The
gauge field $A_\m^a$ as well as the Lagrange multiplier field $b^a$ and
the ghost fields $c^a$, $\cb^a$ belong to the adjoint representation,
whereas the matter fields, collectively denoted by $\f$, belong to some
finite unitary representation of $G$ where the generators will also be
denoted by $T_a$.
\footnote{Fields $\vf$ in the adjoint representation will often be
written as matrices:
\eq
\vf=T_a\vf^a\ .
\eqn{matrix}
The Killing form is taken to be $\d_{ab}$,
\ie the structure constants obey
\eq f_{abc}f^{bcd}=\d_a^d\ .
\eqn{killing}}

The BRS transformations are:
\eq\ba{lcl}
sA_\m \= -\pa_\m c + i[A_\m,c]= -D_\m c   \ ,\\
sc \= -{i\over2} \{c,c\} = - ic^2\ ,\\
s\cb\= b\ ,\\
s\f\= -i c\f\ .
\ea\eqn{brs}
They are nilpotent.

The BRS-invariant gauge fixed classical action in the Landau gauge reads
\eq\ba{l}
\S_{\rm Landau} \\[2mm]
= \intx\lp - {1\over 4g^2}  F_a^{\m\n}F^a_{\m\n}
   + \LL_{\rm matter}(\f,D_\m \f)
   +  b_a\pa^\m A^a_\m +  \cb_a \pa^\m D_\m c^a \rp \ .
\ea\eqn{actionclass}
We don't specify the part of the
gauge invariant action which depends on the matter field and on
its covariant derivative $D_\m\f=(\pa_\m - iA_\m) \f$, only restricted by
the usual power-counting renormalizibility condition.

Let us define
\eq
\ccinq = {1\over40} d_{a}^{\ bc}f_{bde}f_{cfg}\; c^a c^d c^e c^f c^g \ .
\eqn{c-5}
with $d_{abc}$ the completely symmetric invariant tensor of rank 3.
This is a BRS cocycle ($s$-invariant but not of the form
$s(\cdots)$)~\cite{zumino}. The purpose of this and the following section
is to show the finiteness of $\ccinq$ with the help of the ghost equation
governing the dependence of the theory on the ghost $c$. In order to
control the
renormalization properties of $\ccinq$ let us couple it to an external
field $\r$. It will turn out to be useful  to couple also certain
$c$-monomials
of degree 4, 3 and 2 to external fields $\eta$, $\o$, $\tau$ and
$\s$.
Coupling moreover the BRS variations of $A_\m$ and $\f$ to the external
fields $\O^\m$ and $Y$, we write the external field dependent part of
the classical action as:
\eq\ba{rl}
\S_{\rm ext} = \intx \Lp&\!\! \O^{\mu}_a sA_{\m}^a + Y s\f \\
& +\r\ccinq + {1\over8} \eta^{ab} f_{acd}f_{bef} c^c c^d c^e c^f \\
& +\half( \o^{ab}+\tau^{ab})c_a f_{bcd} c^c c^d
  +\half \s^{ab}c_a c_b \Rp\ .
\ea\eqn{action-ext}
with
\eq\ba{lcl}
\eta^{ab} ={\ } \eta^{ba}\ ,\qq \omega^{ab} = {\ }\omega^{ba} \ ,\\
\tau^{ab} = -\tau^{ba}\ ,\qq \sigma^{ab} = -\sigma^{ba} \ .
\ea\eqn{antisymm}

Dimensions and ghost numbers of the different fields are given in
Table~\ref{dim-gh}.

\begin{table}[hbt]
\centering
\begin{tabular}
{|l|| r|      r|   r|     r|   r|        r|    r|       r|     r|       r| }
\hline
    &$A_\m$ &$c$ &$\cb$ &$b$ &$\O^\m$  &$\r$ &$\eta$  &$\o$  &$\tau$  &$\s$
\\ \hline
$d$ &1      &0   &2     &2   &3        &4    &4       &4     &4       &4
\\ \hline
$g$ &0      &1   &$-1$  &0   &$-1$     &$-5$ &$-4$    &$-3$  &$-3$    &$-2$
 \\ \hline
\end{tabular}
\caption[t1]{Dimensions $d$ and ghost numbers $g$.}
\label{dim-gh}
\end{table}

BRS invariance of the external field part of the action together with
nilpotence is achieved by
demanding the following transformation rules ($\O^\m$ and $Y$ being kept
invariant as usual):
\eq\ba{lcl}
s\r = 0\ , \qq                                   \\
s\eta^{ab} = 2\o^{ab}\ ,\qq s\o^{ab} = 0 \ ,\\
s\tau^{ab} = \s^{ab}\ ,\qq s\s^{ab} = 0 \ .
\ea\eqn{brs-ext}

Invariance of the total classical action
\eq
\S(A,c,\cb,b,\O,Y,\r,\eta,\o,\tau,\s)= \S_{\rm Landau}
   + \S_{\rm ext}
\eqn{totalaction}
under the BRS transformations \equ{brs} and \equ{brs-ext}
is expressed by the {\em Slavnov identity}\footnote{The functional
derivative of a functional $\FF$ with respect to a symmetric
or antisymmetric tensor field $t_{ab}$
is defined through the variation formula
\eq
\d\FF = \intx \half \d t_{ab} \fud{\FF}{t_{ab}} \ .
\eqn{funct-der}
}
\eq\ba{rl}
\SS(\S) =&\intx \LP \fud{\S}{\O^\m_a}\fud{\S}{A_\m^a}
     + \half f^{abc} \fud{\S}{\s^{bc}}\fud{\S}{c^a}
     + \fud{\S}{Y}\fud{\S}{\f} +  b^a\fud{\S}{\cb^a} \\
  &{\ }{\ }{\ }{\ }{\ }{\ }{\ }{\ }+ \o^{ab} \fud{\S}{\eta^{ab}}
                + \half\s^{ab} \fud{\S}{\tau^{ab}}\RP
   = 0\ .
\ea\eqn{slavnov}

The action is invariant under the {\em rigid transformations}
$\d_a^{\rm rig}$ of
the group $G$:
\eq
\RR_a^{\rm rig}\S = \sum_\afv \intx\d_a^{\rm rig}\vf\fud{\S}{\vf}=0\ .
\eqn{rigide}
The {\em Landau gauge condition} is expressed by the equation
\eq
\fud{\S}{b_a} = \pa^\m A^a_\m\ .
\eqn{gaugecond}
The {\em antighost equation}
\eq
\bar \GG_a \S = \fud{\S}{\cb^a} + \pa_\m \fud{\S}{\O^a_\m} = 0
\eqn{antighosteq}
follows from the Slavnov identity and the gauge condition.

The {\em ghost equation}, usually valid in the Landau
gauge~\cite{bps-ghost}, extends to the present case as:
\eq
\GG_a \S = \D_a\ ,
\eqn{ghosteq}
where
\eq\ba{rl}
\GG_a = \intx \LP&\!\! \fud{}{c^a} + f_{abc}\cb^b\fud{}{b_c}
   +\half \r d_{abc}\fud{}{\eta_{bc}}                      \\
  &+ \half f_{ab}^{{\ }{\ }d}\eta^{bc}\lp \fud{}{\o^{dc}}
            +\fud{}{\tau^{dc}}\rp                                    \\
  &+ \half \lp (\o_{ab}+\tau_{ab})f^{bcd}
      + 2(\o^{cb}+\tau^{cb})f_{ab}^{{\ }{\ }d} \rp  \fud{}{\s^{cd}} \RP\ ,
\ea\eqn{ghostop}
and
\eq
\D_a = \intx \lp f_{abc}\O^{b\m}A_\m^c + \s_{ab} c^b + iYT_a\f \rp\ .
\eqn{ghostbreak}
Finally it
 is easy to see, using the Jacobi identity obeyed by the structure
constants, that the following {\em ''$\tau$-equation ''} holds:
\eq
\TT_a \S = f_{abc} \fud{\S}{\tau_{bc}} = 0\ .
\eqn{taueq}
The ''linearized'' Slavnov operator
\eq\ba{rl}
\SS_\S =&\intx \LP \fud{\S}{\O^\m_a}\fud{}{A_\m^a}
     + \fud{\S}{A_\m^a}\fud{}{\O^\m_a}
     + \half f^{abc} \fud{\S}{\s^{bc}} \fud{}{c^a}
     + \half f^{abc} \fud{\S}{c^a} \fud{}{\s^{bc}} \\
    &{\ }{\ }{\ }{\ }+ \fud{\S}{Y}\fud{}{\f} + \fud{\S}{\f} \fud{}{Y}
     + b^a\fud{}{\cb^a}
     + \o^{ab} \fud{}{\eta^{ab}}
     + \half\s^{ab} \fud{}{\tau^{ab}}\RP
\ea\eqn{slavnovlin}
is nilpotent if the functional $\S$ is a solution,
not only of the Slavnov
identity \equ{slavnov}, but also of the $\tau$-equation \equ{taueq}:

\eq
\SS(\S) = 0\ ,  {\ }{\ }\TT_a\S = 0{\ }{\ }{\ }{\ }\imply \SS_\S^2=0\ .
\eqn{nilpotency}
%***************************************************************
\subsection{The functional algebra}\label{algebra}
Since BRS nilpotency holds only up to the $\tau$-equation
\equ{taueq}, our first task in the next section will be to prove the
validity of the latter to all orders. We thus will be interested, in the
following, to the space
of functionals $\g$ restricted by:
\eq
\TT_a\g=0
\eqn{taueq2}
In this space the functional operators introduced above obey a nonlinear
algebra whose relevant part reads:
\eq\ba{ll}
(i)&\SS_\g\SS(\g)=0 \ ,                               \\[2mm]
(ii)&\GG_a \SS(\g) + \SS_\g (\GG_a\g -\D_a)= \RR^{\rm rig}_a\g\ , \\[2mm]
(iii)&\fud{}{b^a}\SS(\g) -
      \SS_\g \lp \fud{\g}{b^a}-\pa A_a \rp = \bar\GG_a\g\ ,   \\[2mm]
(iv)&\bar\GG_a\SS(\g)+\SS_\g\bar\GG_a = 0\ ,          \\[2mm]
(v)&\TT_a\SS(\g) = 0 \ ,                               \\[2mm]
(vi)&\fud{}{b^a}(\GG_b\g-\D_b){\ }-{\ }\GG_b(\fud{\g}{b^a}-\pa A_a){\ }=0
                             \ ,                     \\[2mm]
(vii)&\GG_a\bar\GG_b \g + \bar\GG_b(\GG_a\g-\D_a) =
          f_{abc} \lp \fud{\g}{b_c}-\pa A^c \rp  \ ,   \\[2mm]
(viii)&\TT_a(\GG_b\g-\D_b) =0 \ ,                        \\[2mm]
(ix)&\lac \GG_a,\GG_b \rac  = 0\ .
\ea\eqn{algebre}
%*********************************************************************
\section{Renormalization}\label{renormalization}   %3
We want to show that the functional identities of the preceding section
hold to all orders for the vertex functional $\G$ -- which coincides with
the classical action $\S$ \equ{totalaction} in the tree graph approximation.

Let us consider generic classical identities
\eq
\FF_i\S = 0 \ .
\eqn{generic}
According to the quantum action principle~\cite{lam,cl}, the possible
breakings of such identities
by the radiative corrections
are given at their lowest nonvanishing loop order by
local field functionals $\D_i$ with quantum
numbers and dimensions prescribed by those of the corresponding
functional operators:
\eq
\FF_i\G = \D_i + O(\hbar\D) \ .
\eqn{quantumbr}
 The functional operator algebra implies a set of
consistency conditions on the $\D$'s which one has to solve. If the
general solution has the form
\eq
\D_i = \FF_i \hat\D\quad \forall i \ ,
\eqn{triviality}
then $\hat\D$ can be absorbed as a counterterm in the action, and the
identities \equ{generic} are renormalizable\footnote{In \equ{generic}
we have ommitted a
possible classical breaking -- \ie a breaking linear in the quantum
fields like in the ghost equation \equ{ghosteq}.
If the operator $\FF_i$ is nonlinear, it
is its
linearized form (\cf \equ{slavnovlin}) which appears in \equ{triviality}.}.

The gauge fixing condition \equ{gaugecond} and the antighost equation
\equ{antighosteq} are known to be renormalizable~\cite{piguetrouet} and
thus will be assumed to hold:
\eq
\fud{\G}{b^a} = \pa A_a\ ,\qquad \bar \GG_a \G  = 0\ .
\eqn{landau-antigh-ren}

The same can be assumed for the rigid invariance~\cite{brsyug}
(see \equ{rigide}):
\eq
\RR_a^{\rm rig}\G=0
\eqn{rig-ren}
%**********************************************************************
\subsection{The $\tau$-equation}\label{tau-equation}
The most general quantum breaking of the $\tau$-equation \equ{taueq} -- of
dimension 0 and ghost number 3 -- has the form
\eq
\QQ_a =  t_{abcd} c^b c^c c^d \ .
\eqn{tau-br}
Using \equ{killing} we can write it as
\eq
\QQ_a = \half\TT_a \intx f^{def}\tau_{ef}t_{dghl}c^g c^h c^l\ .
\eqn{tautriv}
Hence the $\tau$-equation
\eq
\TT_a \G  = 0\ .
\eqn{taueq-ren}
can be assumed to hold to all orders. As a consequence we shall be able to
exploit the functional algebra displayed at the end of the preceding
section.
%*********************************************************************
\subsection{The ghost equation}\label{ghost-equation}
To discuss the renormalization of the ghost equation \equ{ghosteq} let us
write it as:
\eq
\GG_a \G = \D_a{\ }{\ }+{\ }{\ }\Xi_a  \ ,
\eqn{ghostgam}
where $\Xi_a$ represents the possible breaking induced by the radiative
corrections. $\Xi_a$ is an integrated local functional with dimensions four
and ghost number $-1$ which, according to the nonlinear algebra \equ{algebre},
satisfies the conditions:
\eq
\fud{\Xi_a}{b^c} = 0\ ,\qquad \bar \GG_c \Xi_a  = 0\ ,
\eqn{condit1}
\eq
\RR_a^{\rm rig}\Xi_b = -f_{abc}\Xi^c \ , \qquad \TT_a \Xi_b  = 0\ ,
\eqn{condit2}
and
\eq
\GG_a\Xi_b{\ }{\ }+{\ }{\ }\GG_b\Xi_a = 0\ .
\eqn{condit3}
Equation \equ{condit1} implies that $\Xi_a$ is $b$-independent and depends on
the fields $\cb$ and $\Omega^\mu$ only through the combination
\eq
\gamma_{\mu} = \partial_{\mu}\cb {\ }+{\ }\Omega_{\mu} \ .
\eqn{combination}
It follows then that $\Xi_a$ can be parametrized as
\eq\ba{rl}
\Xi_a = \intx \Lp&\!\! t_{abcde}\rho c^b c^c c^d c^e
       + m_{abcdef} \eta^{bc} c^d c^e c^f  + r_{abcde} \omega^{bc} c^d c^e \\
& + n_{abcde} \tau^{bc} c^d c^e + q_{abcd} \sigma^{bc} c^d
  + p_{abc} \gamma^{b{\mu}} A^c_{\mu} + v YT_a\phi \Rp \ ,
\ea\eqn{Xiexpress0}
where, from \equ{condit2}, $(t, m, r, n, q, p)$ are invariant tensors in the
adjoint representation, and
\eq
f^{mbc} n_{abcde} = 0\ .
\eqn{condit4}
Since
\eq
\intx{\ }p_{abc}\gamma^{b\mu} A^c_\mu{\ }{\ }
         ={\ }{\ }\GG_a \intx{\ }c^m p_{mbc}\gamma^{b\mu} A^c_\mu \ ,
\eqn{triv1}
and
\eq
\intx{\ }YT_a\phi{\ }{\ }={\ }{\ }\GG_a \intx{\ }c^m YT_m\phi \ ,
\eqn{triv2}
it follows that the nontrivial part of $\Xi_a$ reduces to
\eq\ba{rl}
\Xi_a = \intx \Lp&\!\! t_{abcde}\rho c^b c^c c^d c^e
       + m_{abcdef} \eta^{bc} c^d c^e c^f  + r_{abcde} \omega^{bc} c^d c^e \\
& + n_{abcde} \tau^{bc} c^d c^e + q_{abcd} \sigma^{bc} c^d  \Rp \ ,
\ea\eqn{Xiexpress}
i.e. $\Xi_a$ depends only on the variables $(c, \rho, \eta, \omega, \tau,
\sigma)$ and does not contain any space-time derivative. To study the
condition \equ{condit3} on the local functional space \equ{Xiexpress} we
introduce a dimensionless space-time independent parameter $\xi^a$ with zero
ghost number and we consider the operator
\eq
{\DD}{\ }={\ }\xi^a \GG_a  \ ,
\eqn{Doperat}
which, due to the relation $(ix)$ of equation \equ{algebre}, turns out to be
nilpotent:
\eq
{\DD}^2 = 0 \ .
\eqn{Dnilpot}
The introduction of the operator $\DD$ allows to transform the equation
\equ{condit3} into a cohomology problem. Indeed, it is easy to check that if
the general solution of the equation
\eq
{\DD}X = 0 \ ,
\eqn{Dcohom}
where $X$ is an integrated local functional in the variable $(\xi, c, \rho,
\eta, \omega, \tau, \sigma)$ with dimensions 4 and ghost number $-1$, is of the
form
\eq
X{\ }={\ }{\DD}\hat X \ ,
\eqn{trivDcohm}
then the general solution of equation \equ{condit3} reads:
\eq
\Xi_a{\ }={\ }\GG_a \hat\Xi \ ,
\eqn{trivXisol}
which implies the absence of anomalies for the ghost equation \equ{ghosteq}.
The most general form for the $X$-space is given by
\eq\ba{rl}
X{\ } ={\ } \intx \Lp&\!\! L_{abcd}(\xi)\rho c^a c^b c^c c^d
       + R_{abcde}(\xi) \eta^{ab} c^c c^d c^e                 \\
    &+ S_{abcd}(\xi) \omega^{ab} c^c c^d
    + U_{abcd}(\xi) \tau^{ab} c^c c^d + V_{abc}(\xi) \sigma^{ab} c^c  \Rp \ ,
\ea\eqn{Xexpress}
where the invariant tensors $L, R, S, U, V$ are power series in $\xi$.
To characterize the cohomology of $\DD$ we introduce the filtering operator
\eq\ba{rl}
 N &= \xi^a \pad{\ }{\xi^a} {\ }
       +   {\ }\intx \LP c^a\fud{}{c^a} + \rho\fud{}{\rho} \\
   &+ \half\lp\eta^{ab}\fud{}{\eta^{ab}}
                   + \omega^{ab}\fud{}{\omega^{ab}}\rp
  + \half\lp\tau^{ab}\fud{}{\tau^{ab}} + \sigma^{ab}\fud{}{\sigma^{ab}}\rp
\RP \ ,
\ea\eqn{filtration}
according to which, the operator $\DD$ decomposes as
\eq
\DD {\ }={\ } {\DD}^0 + {\DD}^1 \ ,
\eqn{Ddecomp}
where
\eq
{\DD}^{0} {\ }={\ }\intx{\ }\xi^a\fud{}{c^a} \ , \qquad
{\DD}^0 {\DD}^0 = 0 \ .
\eqn{D0express}
It is now apparent that, due to the absence of space-time derivatives in the
general expression \equ{Xexpress}, the cohomology of ${\DD}^0$ vanishes and,
since the cohomology of $\DD$ is isomorphic to a subspace of the
cohomology of ${\DD}^0$~\cite{bandelloni-dixon}, it follows that also $\DD$
has trivial cohomology on $X$.
This proves that the ghost equation \equ{ghosteq} can be implemented to all
orders of perturbation theory.

%*********************************************************************
\subsection{The Slavnov identity}\label{Slavnov identity}

Having renormalized the $\tau$ and the ghost equations, let us discuss the
quantum extension of the Slavnov identity \equ{slavnov}.
{}From the nonlinear algebra \equ{algebre} one has that the breaking of the
Slavnov identity
\eq
\SS(\Gamma) = {\hbar}^n {\D}{\ }+{\ }O({\hbar}^{n+1})  \ ,
\eqn{anslavnov}
has to satisfy the conditions:
\eq
\fud{\D}{b^c} = 0\ ,\qquad \bar \GG_c {\D}  = 0\ ,
\eqn{slavcond1}
\eq
\RR_a^{\rm rig}{\D} = 0 , \qquad \TT_a {\D} = 0\ , \qquad
\GG_a {\D} = 0   \ ,
\eqn{slavcond2}
and
\eq
\SS_{\Sigma}{\D} = 0  \ ,
\eqn{slavcond3}
where $\SS_{\Sigma}$ is the linearized operator defined in \equ{slavnovlin}
and $\D$ is an
integrated local polynomial in the fields with ghost-number 1 and dimension
four.
The index $n$ appearing in \equ{anslavnov} denotes the lowest nonvanishing
order for $\D$.
\par
As in the previous section, the conditions \equ{slavcond1} imply that
$\D$ is $b$-independent and that the fields $\bar c$ and $\Omega_\mu$
enter only in the combination $\gamma_{\mu}$ \equ{combination} so that
$\D$ can be parametrized as:
\eq\ba{l}
{\D}{\ } ={\ } \AA(c,A_\mu,\phi)
           + \intx \Lp \MM_{abcdef}\rho c^a c^b c^c c^d c^e c^f
           + \RR_{abcdefg} \eta^{ab} c^c c^d c^e c^f c^g    \\
\qquad   + \SS_{abcdef} \omega^{ab} c^c c^d c^e c^f
           + \UU_{abcdef} \tau^{ab} c^c c^d c^e c^f
           + \VV_{abcde} \sigma^{ab} c^c c^d c^e            \\
\qquad   + \LL_{abcd} c^a c^b \gamma^{c\mu} A_{\mu}^d
           + \alpha f_{abc} c^a c^b \pa\gamma^c
           + \beta d_{abc} c^a \pa^{\mu} c^{b} \gamma_{\mu}^{c}
           + \lambda f_{{\ }ab}^c c^a c^b YT_c\phi     \Rp \ ,
\ea\eqn{AAexpress}
where $\MM$, $\RR$, $\SS$, $\UU$, $\VV$ and $\LL $
are invariant tensors in the adjoint
representation and $\AA(c,A_{\mu},\phi)$ depends only on the ghost $c$, on the
gauge field $A_{\mu}$ and on the matter fields $\phi$.
It is easy to see that the ghost condition in \equ{slavcond2} implies that
\eq
\MM = \RR = \SS = \UU = \VV = \LL = 0 \ ,
\eqn{MRSUVL}
\eq
\alpha = \beta = \lambda = 0 \ ,
\eqn{abl}
and
\eq
\intx \fud{\AA}{c^a} {\ }={\ }0 \ ,
\eqn{ghostder}
from which it follows that $\AA$ depends only on the space-time derivatives
of $c$.
\par
Finally, the Slavnov condition \equ{slavcond3} reads:
\eq
\intx Tr \Lp -(D_\m c)\fud{}{A_\mu} -ic^2\fud{}{c} - ic\phi \fud{}{\phi}
       \Rp \AA{\ }=0  \ .
\eqn{AAcond}
The most general solution of \equ{AAcond} has been given in~\cite{brsyug} and
coincides, modulo a $\SS_{\Sigma}$ coboundary, with the well known
expression for the gauge anomaly~\cite{abbj}:
\eq
\AA{\ }={\ }\varepsilon^{\mu\nu\rho\sigma}\intx \pa_{\mu}c^a \Lp
  d_{abc}\pa_{\nu}A^b_\rho A^c_\sigma -
  {\DD_{abcd}\over12}A^b_\nu A^c_\rho A^d_\sigma \Rp   \ ,
\eqn{anomaly}
with
\eq
\DD_{abcd} = d_{{\ }ab}^{n}f_{ncd} + d_{{\ }ac}^{n}f_{ndb} +
            d_{{\ }ad}^{n}f_{nbc} \ .
\eqn{DDtensor}
\vspace{5mm}

To summarize this section one has that the vertex functional $\Gamma$
\eq
\Gamma{\ }={\ }\Sigma + O({\hbar})   \ ,
\eqn{summary}
obeys:
\par

i) ${\ }{\ }{\ }$the gauge-fixing condition and the antighost equation
\eq
\fud{\Gamma}{b_a} = \pa A^a   ,\qquad \bar \GG_a {\Gamma}  = 0\ ,
\eqn{summ1}
\par

ii) ${\ }$the rigid gauge invariance and the $\tau$ equation
\eq
\RR_a^{\rm rig}{\Gamma} = 0 , \qquad \TT_a {\Gamma} = 0\ ,
\eqn{summ2}
\par

iii) the ghost equation
\eq
\GG_a {\Gamma} = \Delta_a   \ ,
\eqn{summ3}
\par

iv) the anomalous Slavnov identity
\eq
\SS(\Gamma){\ }={\ }{\hbar}^n r  \AA {\ }{\ }+{\ }{\ }O({\hbar}^{n+1})  \ ,
\eqn{summ4}
where $r$ is a coefficient directly computable in terms of Feynman
diagrams. Moreover, as we shall see in the next sections, the coefficient
$r$ turns out to satisfy a nonrenormalization
theorem~\cite{bardeen,padua,bbbc}.

%*********************************************************************
\section{The Callan-Symanzik equation.}\label{Callan-Symanzik} %4

This section is devoted to the derivation of the Callan-Symanzik equation
obeyed by the vertex functional $\Gamma$.
\par
Equations \equ{summ1} -- \equ{summ4} show that, besides the gauge fixing
condition, the antighost equation and the $\tau$ equation, the functional
$\Gamma$ obeys the ghost equation \equ{summ3}. This equation governs the
dependence of $\Gamma$ on the ghost field $c$ and will impose quite strong
constraints on the Callan-Symanzik equation; it will imply the absence of
anomalous dimensions for the ghost and for the cocycle $\ccinq$ of
equation \equ{c-5}.
\par
Moreover, the presence of the gauge anomaly at the order $\hbar^n$
in the Slavnov identity
\equ{summ4} does not allow for a Callan-Symanzik equation which is
invariant to all orders of perturbation theory:
one expects a Slavnov invariant Callan-Symanzik
equation up to the order
$\hbar^{n-1}$.
\par
However, it will be shown that the Callan-Symanzik equation will extend
to the order $\hbar^n$, which is the order of the
gauge anomaly; this property will be crucial for the nonrenormalization
theorem of the anomaly coefficient $r$ in
\equ{summ4}~\cite{bardeen,padua,bbbc}
\par
To characterize the scaling properties of the model we look for a basis
of insertions which are invariant under the set of equations satisfied
by the functional $\Gamma$: i.e. we look at the most general integrated
local polynomial in the fields ${\hat \Sigma}_{loc}$ with dimension four
and zero ghost number which satisfies the stability
conditions~\cite{bps-ghost,piguetrouet}:
\eq
\fud{{\hat \Sigma}_{loc}}{b^c} = 0\ ,\qquad
\bar \GG_c {{\hat \Sigma}_{loc}}  = 0\ ,
\eqn{stab1}
\eq
\RR_a^{\rm rig}{{\hat \Sigma}_{loc}} = 0\ , \qquad
\TT_a {{\hat \Sigma}_{loc}} = 0\ , \qquad
\GG_a {{\hat \Sigma}_{loc}} = 0   \ ,
\eqn{stab2}
and
\eq
\SS_{\Sigma}{{\hat \Sigma}_{loc}} = 0  \ .
\eqn{stab3}
Proceeding as in the previous sections, it is not difficult to show that
the most general expression for ${\hat \Sigma}_{loc}$ reads:
\eq
{\hat \Sigma}_{loc} {\ }={\ }
  - {z_{g}\over 4g^2}\intx  F_a^{\m\n}F^a_{\m\n} {\ }+
   {\ }\SS_{\Sigma}\intx \Lp z_A \gamma^{a\mu} A_{a\mu}
     - z_{\phi} Y \phi \Rp {\ }+{\ }\sum_i z_i {\hat \Sigma}_i(\phi) \ ,
\eqn{countert}
where $z_{g}$, $z_A$, $z_{\phi}$ and $z_i$ are arbitrary parameters and
the ${\hat \Sigma}_i$ are all possible invariant matter self interactions.
Expression \equ{countert} represents the most general invariant
counterterm. Note that there is no term in the external fields $\r$,
$\eta$, $\o$, $\tau$ and $\s$: this is due to the last condition
\equ{stab2}. Thus one
sees that the conditions \equ{summ1} -- \equ{summ4}
determine the model up to a renormalization of the gauge coupling constant
$g$ (given by $z_g$), of the matter self couplings (given by $z_i$) and to
a redefinition of the gauge field amplitude ($z_A$) and of the matter
fields ($z_{\phi}$).
\par
Expression \equ{countert} can be rewritten as
\eq
{\hat \Sigma}_{loc} = - z_g g^2 {\partial\Sigma \over \partial g^2}{\ }+{\ }
 z_A \NN_A\Sigma {\ }+{\ }z_\phi \NN_{\phi}\Sigma
 {\ }+{\ }\sum_i z_i \lambda_i {\partial \Sigma \over \partial \lambda_i} \ ,
\eqn{parametric}
where $\Sigma$ is the classical action \equ{totalaction} and $\NN_A$,
$\NN_{\phi}$ are the Slavnov invariant counting operators defined by
\eq
\NN_A = \intx \Lp A^{a\mu}\fud{}{A^{a\mu}} - b^a\fud{}{b^a}
        -{\bar c}^a\fud{}{{\bar c}^a} -
          \Omega^{a\mu}\fud{}{\Omega^{a\mu}} \Rp  \ ,
\eqn{counting1}
\par
\eq
\NN_{\phi} = \intx \Lp \phi\fud{}{\phi} - Y\fud{}{Y}   \Rp \ .
\eqn{counting2}
and $\lambda_i$ are the matter self coupling constants.
\par
It is apparent from
\equ{parametric} that the set
\eq
\Lp {\partial \Gamma \over \partial g},
\qquad {\partial \Gamma \over \partial \lambda_i}, \qquad \NN_A\Gamma,
 \qquad \NN_{\phi}\Gamma \Rp \ ,
\eqn{set}
can be taken as a basis for the insertions of dimensions four and
ghost number zero which are Slavnov invariant and which are compatible with
the gauge fixing condition and the antighost equation \equ{summ1} as well
as with the $\tau$ and the ghost equations \equ{summ2} -- \equ{summ3}.

The insertion defined by $\partial\G/\partial\m$, where $\m$ is the
normalization point,
obeys to the same conditions. Expanding it in the basis \equ{set}
we get then  the Callan-Symanzik equation
\eq\ba{rl}
\CC\Gamma{\ }=& {\ }\Lp \mu \pad{\ }{\mu} +
    \hbar \beta_g {\partial {\ }\over \partial g}
   + \hbar \sum_i \beta_i {\partial {\ }\over \partial \lambda_i}
   + \hbar \gamma_A \NN_A
   +\hbar \gamma_{\phi} \NN_{\phi} \Rp \Gamma{\ }                 \\
         =&{\ }\hbar^n \Delta^n_c   + O(\hbar^{n+1})   \ ,
\ea\eqn{calla-sym}
where $\Delta^n_c$ is an integrated local polynomial corresponding to the
lack of the Slavnov invariance at the order $\hbar^n$ according to equation
\equ{summ4}.
\par
Notice that the validity of the ghost equation \equ{summ3} has led to
the absence of anomalous dimensions for the ghost field $c$.
\par
{}From the algebraic relation
\eq
\CC\SS(\Gamma){\ }={\ }\SS_{\Gamma}\CC\Gamma  \ ,
\eqn{relalg}
it follows that, applying the Callan-Symanzik operator to the anomalous
Slavnov identity \equ{summ4} and retaining the lowest order in $\hbar$,
one gets the equation:
\eq
\SS_{\Sigma} \Delta^n_c{\ }={\ }\mu {\partial r \over \partial \mu}\AA \ ,
\eqn{muequ}
which, due to the fact that $r$ is a dimensionless coefficient, implies that
\eq
\SS_{\Sigma}\Delta^n_c = 0 \ .
\eqn{deltacond}
This condition tells us that the local breaking $\Delta^n_c$, in spite
of the presence of the gauge anomaly, is Slavnov invariant and that it can be
expanded in the basis \equ{set}.
This allows the  extension of
the Callan-Symanzik equation in a Slavnov invariant
way up to the order $\hbar^n$, i.e.:
\eq
\CC\Gamma{\ }={\ }\hbar^{n+1} \Delta^{n+1}_c {\ }+{\ }O(\hbar^{n+2}) \ ,
\eqn{callanext}
with $\Delta^{n+1}_c$ local.
\par
Deriving this equation with respect to the source $\rho(x)$ we get
\eq
\CC[\ccinq \cdot \Gamma]{\ }= O(\hbar^{n+1}) \ ,
\eqn{c5finit}
which expresses the finiteness, \ie the vanishing of its anomalous
dimension,  of the ghost cocycle \equ{c-5}, whose quantum extension is
defined by
\eq
[\ccinq \cdot \Gamma]=\fud{\G}{\rho}\ .
\eqn{ccinqdef}
%**********************************************************************
\section{The descent equations}\label{sect-desc} %5
The purpose of this section is to show that the gauge anomaly
operator \equ{anomaly}, here written in terms of differential
forms\footnote{The symbole $\tr$ means the normalized trace
\eq
\tr (T_a T_b) = \d_{ab}\ .
\eqn{tracenorm}
The symmetric tensor $d_{abc}$ in \equ{c-5} is defined by
\eq
d_{abc} = \half \tr\lp T_a\{T_b,T_c\}\rp \ .
\eqn{d-abc}
$\o^q_p$ indicates a $p$-form of ghost number $q$,
the ghost
number of $c$ being equal to 1. The gauge 1-form is $A=A_\m dx^\m$, its
BRS transformation reads
\eq
sA = dc - i[c,A]\ .
\eqn{brs-form}
By convention $d\o = dx^\m\pa_\m\o$ for any form $\o$, and $s$
anticommutes with the exterior derivative $d$. Brackets $[\ ]$ denote
graded commutators.}:
\eq\ba{l}
\AA(A,c)= \int \o^1_4 \ ,\\[2mm]
\o^1_4 = \half \tr\lp dc( AdA + dAA -iA^3) \rp  \ ,
\ea\eqn{anomaly-form}
can be renormalized in such a way that it obeys a Callan-Symanzik equation
with vanishing anomalous dimension. This will follow from the vanishing
of the anomalous dimension of the cocycle insertion \equ{ccinqdef} proved in
Section~\ref{Callan-Symanzik}.

Throughout this section we shall assume the validity of the gauge
condition and antighost equation \equ{summ1}, and of the rigid invariance.
This means that we shall deal with functionals of $A$, $\f$, $c$, $\g$
(see \equ{combination}), $Y$ and $\s$ which is the external
field coupled to the BRS variation of $c$, see Eq. \equ{action-u} below.
%*********************************************************************
\subsection{Renormalization of the anomaly operator}\label{renanomalie}
The starting point is the relation of
the ghost cocycle \equ{c-5} to the gauge anomaly
through the descent equations~\cite{zumino}
\eq\ba{l}
s\om{q}{5-q} + d\o^{q+1}_{4-q} = 0\ ,\qquad q = 1,\cdots, 4 \ ,\es
s\o^5_0 = 0\ .
\ea\eqn{descent-eq}
where
\eq\ba{l}
\om{2}{3}= - \tr \lp (dc)^2 A\rp    \ ,\es
\om{3}{2}=  \tr\lp (dc)^2 c\rp  \ ,\es
\om{4}{1}=  -{i\over2} \tr \lp dc\,c^3\rp  \ ,\es
\om{5}{0}=  -{1\over10} \tr\lp c^5\rp  \ .
\ea\eqn{omegas}
We have  rewritten $\ccinq$ in the matrix
notation as
\eq
\ccinq = \om{5}{0}     \ .
\eqn{c-5form}

What we need is a {\em ladder} of operator
insertions $\{\Om{q}{5-q}\}$ which is a quantum
extension of the classical ladder $\{\om{q}{5-q}\}$ and obeys the quantum
descent equations -- valid to the same order as the Slavnov
identity \equ{summ4} since the nilpotency of $\SS_\G$ is broken
by the anomaly:
\eq\ba{l}
\SS_\G\Om{q}{5-q} + d\Om{q+1}{4-q} =  O(\hbar^n)\ ,
\qquad q = 1,\cdots, 4 \ ,\es
\SS_\G\Om{5}{0} = O(\hbar^n)\ .
\ea\eqn{descent-ren}
The renormalized anomaly operator insertion will be defined as
\eq
\AA_R = \int \Om{1}{4}
\eqn{ren-anom}
and by
\eq
\Om{5}{0} = [\ccinq\cdot\G]
\eqn{om5-0}
which is the finite insertion constructed in
Section~\ref{Callan-Symanzik}.

In order to construct the renormalized insertions $\Om{q}{5-q}$ we
introduce external fields $\uu{-q}{q-1}$, with $\uu{-q}{q-1}$ a
$(q-1)$-form of ghost number
$-q$, singlet of the gauge group~\cite{lps}. We therefore start now
with the classical action
\eq\ba{rl}
\S = \S_{\rm Landau} +  \intx \Lp&\!\! \tr(\g^\mu sA_\m + Y s\f  +\s sc)\\
 &+ \sum_{q=1}^5 \uu{-q}{q-1}\om{q}{5-q}       \Rp\ ,
\ea\eqn{action-u}
with $\S_{\rm Landau}$ given by \equ{actionclass}. We have discarded the
external fields used in Section~\ref{section2} for controling the
finiteness of $\ccinq$, but we have
introduced the external field $\s$ coupled to the BRS variation of $c$.
This action is BRS-invariant, the external fields $u$ transforming as:
\eq\ba{l}
s\uu{-1}{0} = 0\ ,\es
s\uu{-q}{q-1} = -d\uu{-q+1}{q-2} \ ,\qquad q=2,\cdots,5\ .
\ea\eqn{brs-u}
The classical Slavnov identity now reads
\eq\ba{rl}
\SS(\S) = \intx \LP&\!\! \tr\lp\fud{\S}{\O^\m}\fud{\S}{A_\m}
     +  \fud{\S}{\s}\fud{\S}{c}+  b\fud{\S}{\cb}\rp
     + \fud{\S}{Y}\fud{\S}{\f}  \\
  &+ \sum_{q=2}^5 s\uu{q-1}{-q}\fud{\S}{\uu{q-1}{-q}}      \RP
   = 0\ ,
\ea\eqn{slavnov-u}
with a corresponding nilpotent linearized form (\cf~\equ{slavnovlin}).
The rigid Ward identity \equ{rigide}, the gauge condition
\equ{gaugecond} and the antighost equation \equ{antighosteq} are left
unchanged and will again be assumed to hold to all orders for the
vertex functional $\G$.

The construction of a  ladder of insertions
fulfilling the quantum descent equations \equ{descent-ren} will be
achieved by showing the validity of the
Slavnov identity \equ{slavnov-u} (anomalous at the order $n$):
\eq
\SS(\G) = {\ }{\hbar}^n r  \AA {\ }{\ }+{\ }{\ }O({\hbar}^{n+1})\ .
\eqn{slavnov-u-ren}
We have thus to show the triviality of
the cohomology of $\sx$ in the sector of ghost number 1
for the $u$-dependent integrated local functionals
\eq
\D =\int \ch{1}{4} \ .
\eqn{delt}

The condition $\sx\D=0$ and the triviality~\cite{bonora,brandt}
of the cohomology of $d$ mean
that the 4-form $\ch{1}{4}$ belongs to a ladder of
forms obeying descent equations similar to \equ{descent-eq}:
\eq\ba{l}
\sx\ch{q}{5-q} + d\chi^{q+1}_{4-q} = 0\ ,\qquad q = 1,\cdots, 4 \ ,\es
\sx\chi^5_0 = 0\ .
\ea\eqn{descent-eq-chi}
We have first to solve the last of these equations, which is a problem of
{\em local} cohomology (see Def. A.2 in Appendix A).
Then we have to solve the remaining
equations, for increasing form degree, which again is a problem of local
cohomology.

In the absence of the external fields $u$ the local
cohomology depends only on the invariants $\tr c^n$ and on those
made with the Yang-Mills curvature $F$ and its
der\-iv\-at\-ives\cite{bandelloni-dixon}. This uniquely leads to the
ladder \equ{descent-eq} and to the chiral anomaly \equ{anomaly-form}.

It is shown in the Appendix that the
dependence of the cohomology on the external fields $u$ occurs only
through the zero-form $\uu{-1}{0}$. The most general $u$-dependent
expression for $\ch{5}{0}$ in the local cohomology
would be the superposition of the monomials $\uu{-1}{0}\tr c^6$ and
$\uu{-1}{0}(\tr c^3)^2$, but they vanish due to
the cyclicity of the trace and the anticommutativity of the ghost fields.
This proves the absence of $u$-dependent anomalies, hence of
any obstruction to the construction of the renormalized ladder
\equ{descent-ren}.
%************************************************************************
\subsection{Callan-Symanzik equation for the anomaly}\label{csanomalie}
In order to prove the vanishing of the anomalous dimension for
the renormalized anomaly operator defined in the preceding subsection
(see \equ{ren-anom}), we have to repeat in presence of the $u$-fields
the construction of the Callan-Symanzik equation
performed in Section~\ref{Callan-Symanzik}, and to use the fact that the
$\ccinq$ insertion \equ{om5-0} has no anomalous dimension.

The basis of classical invariant insertions used in Eq.~\equ{countert}
must be completed by adding to it all BRS invariants of the form
\eq
\D =\int \ch{0}{4} \ ,
\eqn{delta-ct}
where $\ch{0}{4}$ is a 4-form of zero ghost number, depending on
the $u$'s. Like in the case Eqs.~\equ{delt} and \equ{descent-eq-chi},
the latter belongs to a ladder obeying the
descent equations:
\eq\ba{l}
\sx\ch{q}{4-q} + d\chi^{q+1}_{3-q} = 0\ ,\qquad q = 0,\cdots, 3 \ ,\es
\sx\chi^4_0 = 0\ .
\ea\eqn{descent-ct}
The most general nontrivial expression for $\ch{4}{0}$
depending only on $\uu{-1}{0}$ and on $c$ according to the proposition
of Appendix A, is $\uu{-1}{0}\tr c^5$. It leads to the expression, unique
modulo $\SS_\S$:
\eq
\int \ch{0}{4} = \int  \sum_{q=1}^5 \uu{-q}{q-1}\om{q}{5-q}
 = \NN_u \S    \ ,
\eqn{ucounterterm}
with the $u$-counting operator defined by:
\eq
\NN_u = \int \sum_{q=1}^5 \uu{-q}{q-1} \fud{}{\uu{-q}{q-1}}\ .
\eqn{u-counting}
The second equality in \equ{ucounterterm} follows from the observation
that its left-hand side consists just of the $u$-dependent terms
of the action \equ{action-u}.
The most general form for the classical invariant $\D$ thus is a linear
superposition of
\equ{ucounterterm} and of the $\SS_\S$-variation of an arbitrary local
field polynomial. The quantum basis \equ{set} can accordingly be completed
by adding to it the insertion
$\NN_u\G$ and the $\SS_\G$-variation of an arbitrary
insertion of ghost number $-1$. This implies that the
Callan-Symanzik equation in presence of the fields $u$ will take the
form
\eq
\CC_{(u)}\Gamma{\ }= {\ } \lp \CC +\hbar \g_u \NN_u \rp\G
     \ =\ \hbar\SS_\G [\hat\D\cdot\G]  + O(\hbar^{n+1})   \ ,
\eqn{cs-u}
where $\hat\D$ is some $u$-dependent insertion of ghost number
$-1$; $\CC$ is the Callan-Symanzik operator defined in \equ{calla-sym}.
The fact that the anomaly starts to produce effects only at the
order $n+1$ follows from the same argument used in
Section~\ref{Callan-Symanzik}.

Differentiation of \equ{cs-u}
with respect to $\uu{-5}{4}$ and use of the anticommutativity of this
derivative with the linearized Slavnov operator $\SS_\G$,
yields the Callan-Symanzik equation  for
the insertion of the cocycle $\ccinq$ with anomalous dimension $\g_u$:
\eq
\lp\CC + \hbar \g_u\rp [\ccinq\cdot\G]
  = -\hbar\SS_\G \lp\fud{}{\uu{-5}{4}}[\hat\D\cdot\G]\rp
                              + O(\hbar^{n+1})= O(\hbar^{n+1})\ .
\eqn{cs-c5}
The last equality follows from the fact that the most general
expression for the $\uu{-5}{4}$-dependent part of $\hat\D$ reads
$\uu{-5}{4}\tr c^4$, which identically vanishes.
But since the insertion
$\ccinq$ was shown to have zero anomalous dimension,
we conclude that
\eq
\g_u=0\ .
\eqn{zeroandim}
Differentiating now \equ{cs-u} with respect to $\uu{-1}{0}$ and
integrating over space-time we get the Callan-Symanzik equation
\eq
\CC\AA_R = -\hbar\SS_\G \int \fud{}{\uu{-1}{0}}[\hat\D\cdot\G]
                                        + O(\hbar^{n+1})\ ,
\eqn{cs-anom}
for the anomaly insertion, without anomalous dimension as announced,
up to an irrelevant $\SS_\G$-variation following from the right-hand
side of \equ {cs-u} and from
the anticommutativity of $\int\d/\d\uu{-1}{0}$ with $\SS_\G$.
%************************************************************************
\section{The non\-ren\-or\-mal\-iza\-tion theo\-rem
                                       of the gau\-ge an\-om\-aly}   %6
As an application of the finiteness properties displayed by the quantum
insertions $\{\Om{q}{5-q}\}$ of \equ{descent-ren}, let us discuss the
nonrenormalization theorem of the gauge anomaly.

Following~\cite{bbbc}, we can extend the anomalous Slavnov
identity \equ{summ4} to the order $\hbar^{n+1}$ as:
\eq
\SS(\Gamma){\ }={\ }{\hbar}^n r  \AA_R {\ }{\ }+{\ }{\ }
 {\hbar}^{n+1}\BB{\ }{\ }+O({\hbar}^{n+2})  \ ,
\eqn{extsl}
where $\AA_R$ is the renormalized anomaly operator insertion defined
in \equ{ren-anom} and $\BB$ is an integrated local functional of
ultraviolet dimension four and ghost number one.

Applying the Callan-Symanzik operator to both sides of equation
\equ{extsl} and using the Callan-Symanzik equation
without anomalous dimension \equ{cs-anom} for $\AA_R$
one gets, to lowest order (\ie order $n+1$) in $\hbar$, the equation:
\eq
\Lp \beta_g^{(1)} \pad{r}{g}{\ }
   +{\ }\sum_i \beta_i^{(1)}\pad{r}{\lambda_i} \Rp \AA{\ }={\ }
   \mu \pad{\BB}{\mu}{\ }+{\ }\SS_{\Sigma}{\bar \Delta}  \ ,
\eqn{nonren1}
where $\AA$ is the gauge anomaly expression \equ{anomaly-form}, $\beta_g^{(1)}$
and $\beta_i^{(1)}$ are respectively the one-loop beta functions
for the gauge and the self matter couplings and $\bar \Delta$ is a
local integrated functional.

Taking into account that $\BB$ is homogeneous of degree zero in the mass
parameters~\cite{bbbc}, i.e.:
\eq
\mu\pad{\BB}{\mu} = 0   \ ,
\eqn{BBind}
and that the gauge anomaly $\AA$ cannot be written as a local
$\SS_{\Sigma}$-variation one has the condition:
\eq
 \beta_g^{(1)} \pad{r}{g}{\ }
   +{\ }\sum_i \beta_i^{(1)}\pad{r}{\lambda_i} {\ }={\ }0  \ ,
\eqn{nonren}
which, in the generic case
$\beta_g^{(1)} \not= 0$ and $\beta_i^{(1)} \not= 0$,
implies~\cite{bbbc} that if the coefficient $r$ vanishes at one loop order it
will vanish to all orders.
%*********************************************************************
\vspace{1cm}

\noindent{\large{\bf Acknowledgments}}: We wish to thank C. Becchi,
A. Blasi, R. Collina and R. Stora
for useful discussions.

%*********************************************************************
\appendix
\newtheorem{defin}{Definition}[section]
\newtheorem{prop}{Proposition}[section]
\section{Local cohomology for complete ladder fields}
Let $\VV$ be the space of nonintegrated local field
functionals, the fields being defined on some differential manifold, and
let us assume a coboundary operator $\d$ acting on $\VV$,
nilpotent and anticommuting with the exterior derivative $d$.
\begin{defin}
A ''complete  ladder field'' on a $D$-dimensional
differentiable manifold  is a set
$\LL^{(Q)} = \{ \uu{Q-p}{p}|p=0,\cdots,D\}$
of $D+1$ differentiable
forms\footnote{$\uu{q}{p}$ is a form of degree $p$ and
ghost number $q$.}, where $Q$ is
some fixed algebraic integer.
The coboundary operator $\d$ acts on the $u$'s as:
\eq\ba{l}
\d\uu{Q}{0} = 0\ ,\es
\d\uu{Q-p}{p} = -d\uu{Q-p+1}{p-1} \ ,\qquad p=1,\cdots,D\ .
\ea\eqn{brs-ladder}
\end{defin}
\begin{defin}
The ''local cohomology'' space is the set of equivalence classes mod\-ulo--$\d$
of solutions $\QQ\in\VV$ of the equation
\eq
\d\QQ = 0\ .
\eqn{cocycleeq}
\end{defin}
\begin{prop}
The local cohomology space depends on the complete
ladder field $\LL^{(Q)}$ only through its element $\uu{Q}{0}$, not derivated.
\end{prop}
\noindent
The proof is a slight generalization of the one given in
Ref.~\cite{brandt}.
The space of local fonctionals we have to consider consists of all the
polynomials in the symmetric and antisymmetric derivatives
of the form components $u^{Q-p}_{[\m_1\cdots\m_p]}$:
\eq\ba{l}
S^{Q-p}_{\m_1\cdots\m_p, \n_1\cdots\n_r} =
          \pa_{\n_1}\cdots\pa_{(\n_r} u^{Q-p}_{\m_1)\cdots\m_p}\ ,
        \quad p=1,\cdots D;\ r\ge0\ ,  \es
A^{Q-p}_{\m_1\cdots\m_p, \n_1\cdots\n_r} =
          \pa_{\n_1}\cdots\pa_{[\n_r} u^{Q-p}_{\m_1]\cdots\m_p}\ ,
        \quad p=1,\cdots D;\ r\ge0\ , \es
A^Q_{\n_1\cdots\n_r} = \pa_{\n_1}\cdots\pa_{\n_r} u^Q\ ,
        \quad  r\ge1\ ,
\ea\eqn{derivatives}
The action of $\d$ on these derivatives reads
\eq\ba{l}
\d S^{Q-p}_{\m_1\cdots\m_p, \n_1\n_2\cdots\n_r} =
                      A^{Q-p+1}_{\m_1\cdots\m_{p-1},\m_p\n_1 \cdots\n_r}\ ,
    \qquad p=1,\cdots D; \ r\ge 0\ ,    \es
\d A^{Q-p}_{\m_1\cdots\m_p, \n_1\cdots\n_r} = 0\ ,
    \qquad p=1,\cdots D;\ r\ge 0\ ,  \es
\d u^Q = 0\ .
\ea\eqn{deltader}
One sees that all fields and their derivatives -- {\em except} the
undifferentiated 0-form $u^Q$ -- are displayed in BRS
doublets, \ie in couples $(U,\ V=\d U)$. It is
known~\cite{brandt,piguetsibold86}
that the cohomology cannot depend on such couples. This ends the proof
of the proposition.
%********************************************************************

%*********************************************************************
\end{document}